\documentclass[ aip, rsi, amsmath,amssymb, reprint, ]{revtex4-1}
\usepackage{graphicx}
\usepackage{dcolumn}
\usepackage{bm}
\usepackage{amssymb}
\usepackage{amsmath}
\usepackage{amsfonts}

\usepackage{hyperref}
\usepackage{bbm}
\usepackage[usenames]{color}
\begin{document}

\preprint{AIP/123-QED}

\title{Dissipative solitons for bistable delayed-feedback systems}

\author{Vladimir V. Semenov}
\email{semenov.v.v.ssu@gmail.com}
\affiliation{Department of Physics, Saratov State University, Astrakhanskaya str. 83, 410012 Saratov, Russia}
\author{Yuri L. Maistrenko}
\email{y.maistrenko@biomed.kiev.ua}
\affiliation{Institute of Mathematics and Centre for Medical and Biotechnical Research, NAS of Ukraine, Tereshchenkivska St. 3, 01030 Kyiv, Ukraine}
\affiliation{Forschungszentrum J{\"u}lich, Central Institute of Engineering, Electronics and Analytics, Electronic Systems (ZEA-2), 52428 J{\"u}lich, Germany}

\date{\today}

\begin{abstract}
We study how nonlinear delayed-feedback in the Ikeda model can induce solitary impulses, i.e. dissipative solitons. The states are clearly identified in a virtual space-time representation of the equations with delay, and we find that conditions for their appearance is bistability of a nonlinear function and negative character of the delayed feedback. Both dark and bright solitons are identified in numerical simulations and physical electronic experiment showing an excellent qualitative correspondence and proving thereby the robustness of the phenomenon. Along with single spiking solitons, a variety of compound soliton-based structures is obtained in a wide parameter region on the route from the regular dynamics (two quiescent states) to developed spatio-temporal chaos. The number of coexisting soliton-based states is fast growing with delay, which can open new perspectives in the context of information storage. 
\end{abstract}

\maketitle

\begin{quotation}
Solitons are solitary waves or wave packets travelling in space. These structures, first reported by J.Russel in 1834,  can be found in many physical, biological, chemical, and other spatially-extended systems. One can distinguish solitons observed in conservative and dissipative systems. The second ones are called {\it \textbf{dissipative solitons}}. They are characterised by structural robustness and can persist for a long time of observation despite of  dissipation due to the presence of a source of energy in an active propagation medium. Surprisingly, as it was found in the last decade,  stable localized patterns topologically equivalent to dissipative solitons can arise in a virtual space of the purely temporal dynamics of systems with delay. In the current paper we reveal the apperance of dissipative solitons in a bistable Ikeda-type system with delay. We report multiple coexistence of bright and dark solitons, from just a singe one to any number as allowed by the system size.  The phenomenon is observed in a wide parameter region at the transition from quiescence to developed spatio-temporal chaos in an excellent qualitative correspondence between numerical simulation and experiment.
\end{quotation}
\section{Introduction}
Dissipative solitons are stable localized structures, which are realized in nonlinear dissipative spatially extended systems \cite{fauve1990,kerner1994,christov1995,gustave2015,akhmediev2008,purwins2010,liehr2013}. Appearance of the dissipative solitons results from the energy supply-dissipation balance and the simultaneous impact of nonlinearity and dispersion. They occur  in optics and optoelectronics \cite{boardman2001,haudin2011,grelu2012,verschueren2013,garbin2015,marconi2015,gustave2015,javaloyes2017}, magnetoelectronics \cite{wong2010,grishin2014}, plasma physics \cite{tur1992,ghost2014,sultana2015}, biology \cite{lautrup2011,akhmediev2013} and chemistry \cite{kai1994,kai1995}. In optics these structures are now under intensive study due to perspectives of practical applications in  optical data processing and communication \cite{hasegawa1995,marin-palomo2017} in the context of development of new generation computing systems \cite{adamatzky2002} such as neuromorphic and reservoir computers \cite{larger2012,paquot2012}. 

There is a strong correspondence between the behaviour of time-delayed systems and the dynamics of ensembles of coupled oscillators or spatially extended systems. Using a virtual space-time representation one can track down spatio-temporal phenomena in the purely temporal dynamics of the time-delayed systems \cite{arecchi1992,giacomelli1996}. This approach consists in interpreting of the delay interval $[0,\tau]$ as a spatial coordinate, then the further dynamics is mapped on space.  Besides the theoretical importance, this tight correspondence between the spatially-extended systems and the time-delayed oscillators opens new vision on possible application of time-delayed systems in transient computing \cite{martinenghi2012,larger2017} and  neuromorphic memory \cite{romeira2016}. 

Examples of dissipative soliton patterns for delayed-feedback oscillators, in both theory and experiments, were  reported recently in optical systems \cite{gustave2015,romeira2016,yanchuk2017,brunner2017}. Mathematical aspects of the dissipative soliton appearance were derived for a reduced time-delayed model of mode-locked laser  \cite{vladimirov2005,nizette2006,garbin2015}, the complex Ginzburg-Landau equation including feedback terms \cite{garbin2015,marconi2015,puzyrev2016}, and the FitzHugh-Nagumo model with time-delay \cite{romeira2016}. 

In the current study we develop a novel approach to dissipative soliton modelling based on a second order modification of the Ikeda time-delayed equation. The proposed model exhibits single as well as multiple dissipative soliton patterns clearly  illuminated in the respective virtual space, which is caused by bistability of the nonlinear function applied through negative delayed feedback.  Such system configuration is different from those proposed earlier in Refs.\cite{larger2013,larger2015}, where the feedback is positive, provoking thereafter the chimera-like behaviour.  Our model does not show chimeras. Instead, we find that single and multiple dissipative solitons robustly develop in this case,  as it is controlled by the strength of nonlinearity. In the study, we combine extensive numerical simulation with physical electronic experiment showing an excellent qualitative correspondence in the soliton appearance.  

The model is symmetric with a bistable nonlinear function. As a result, two soliton types - dark (spiking down) and bright (spiking up) - naturally arise as symmetric copies of each other. We analyze the influence of asymmetry inevitably presenting in experiment and show their robustness at symmetry breaking.  Both dark and bright solitons are obtained in the experiment despite of fluctuations and asymmetry of the setup  (dark one, however, appears to be much more probable due to some inner setup characteristic).  On the other hand, some of more complex soliton-based structures, e.g. combinations of two different quiescent states are found to exist only in the purely symmetric model. They disappear with asymmetry demonstrating the phenomenon of coarsening \cite{giacomelli2012,yanchuk2017}.

\section{Dissipative solitons}
\subsection{The model}
%
\begin{figure}[t!]
\centering
\includegraphics[width=0.45\textwidth]{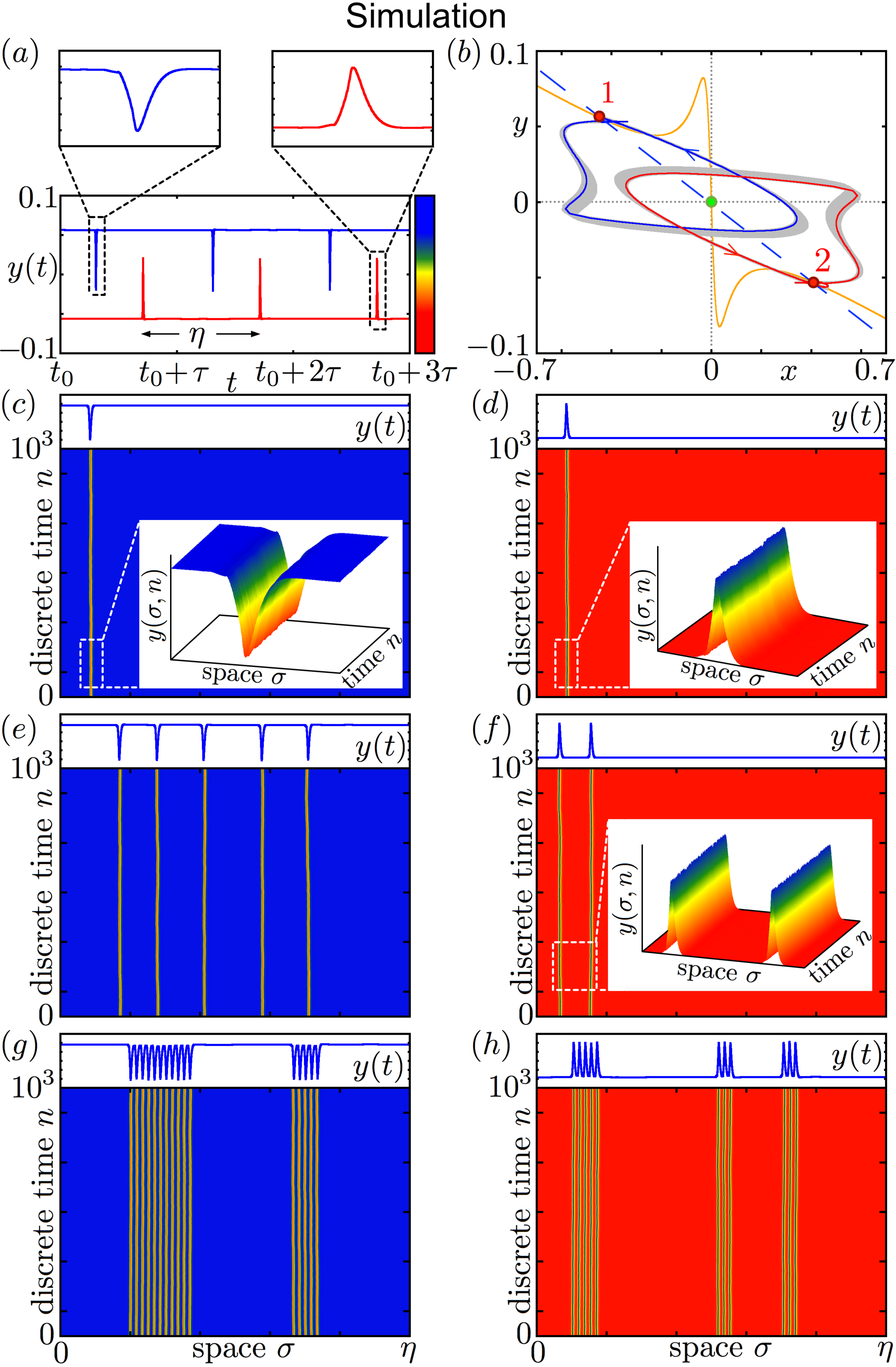} 
\caption{Dissipative solitons in delay dynamics of model (\ref{system}). (a) Two realizations of $y(t)$ in time range $t\in[t_{0}:t_{0}+3\tau]$. (b) Projection of trajectories corresponding to realizations shown in panel (a) on phase plane ($x,y$). Blue and red contours correspond to the realization in one delay range $t\in[t_{0}:t_{0}+\eta]$, grey background shows realizations at $t\in[0:5\times 10^{4}]$, both steady states 1 and 2 are coloured in red, the saddle fixed point in the origin is shown in green, the orange solid line shows the nullcline $\dot{x}=0$, the blue dashed line shows the nullcline $\dot{y}=0$. (c)-(h) Space-time plots of $y(t)$ corresponding to:  (c) one dark soliton (blue trajectory in panels (a),(b)); (d) one bright soliton (red trajectory in panels (a),(b)); (e) five dark solitons; (f) two bright solitons; (g) two bursts of totally sixteen dark solitons; (h) three bursts of totally eleven bright solitons. Parameters are $\varepsilon=0.005$, $a=200$, $b=0.2$, $g=0.1$, $m=8.05$, $\tau=50$. Quasi-space parameter $\eta$ sets to: 
$50.2333$ (panels (c) and (d)), 
$50.2336$ (panel (e)), 
$50.2334$ (panel (f)), 
$50.2323$ (panel (g)), 
$50.2322$ (panel (h)).
All states are obtained from random initial conditions. Transients of $n=10^{6}$ time units were discarded before plotting.}
\label{fig1}
\end{figure}  

Our model is a paradigmatic Ikeda-type oscillator in the form
\begin{equation}
\label{system}
\left\lbrace
\begin{array}{l}
\varepsilon \dot{x} = -y-gx + f[x(t-\tau)], \\
\dot{y}=x+my
\end{array}
\right.
\end{equation}
with bimodal negative delayed-feedback function $f$ chosen, for definiteness, as
\begin{equation}
\label{nonlinearity}
f[x]=-\dfrac{x}{ax^2+b},
\end{equation}
where small parameter $\varepsilon>0$ separates slow and fast motions. Other parameters are $g,a,b,m>0$.  Eqs.~(\ref{system}) were simulated numerically by the Heun method \cite{manella2002} with the time step $\Delta t = 0.0001$ and randomly chosen initial conditions. Parameters were fixed as $\varepsilon=0.005$, $a=200$, $b=0.2$, $g=0.1$, $\tau=50$, $m\in [6.0:8.5]$. Experimental realization of the model (see the next subchapter) shows an excellent qualitative correspondence with simulations.  Similar model with negative delayed feedback and bimodal nonlinearity was recently proposed and studied in the context of the chimera control by periodic and stochastic forcing \cite{semenov2016}. The goal of this study is to give evidence and describe the properties of the fascinating objects, dissipative solitons, multiply and robustly arising in a wide region of the system parameters.

The delay-free dynamics of the model  (i.e. when $\tau=0$ in Eqs.(\ref{system})) is bistable and exhibits the coexistence of two self-oscillatory regimes \cite{semenov2018}.   When the delay $\tau$ is introduced  and it is much larger than the characteristic response time and much smaller than the  time of observation, the system (\ref{system}) exhibits periodically alternating  impulses, {\it dissipative solitons} in some range of the parameters.  Depending on the initial conditions, two kinds of the impulses can be obtained,  see the blue (spiking down) and red (spiking up) realizations in Fig. \ref{fig1} (a).  Two-dimensional images of the impulses in the phase space ($x$,$y$) represent closed loops (periodic orbits) coloured in blue and red respectively [Fig. \ref{fig1} (b)].  Note that most of the time red and blue soliton trajectories spend  in a vicinity of one of the steady states (see equilibrium points 1 and 2 in Fig.~\ref{fig1}~(b)).

Following the standard procedure \cite{arecchi1992,giacomelli1996}, delayed-feedback system  (\ref{system}) is analyzed in a virtual, spatially extended representation. The purely temporal dynamics is mapped onto space-time ($\sigma,n$) by introducing $t=n\eta+\sigma$ with an integer (slow) time variable $n$, and a pseudo-space variable $\sigma \in [0,\eta]$, where $\eta=\tau+\delta$ with a quantity $\delta$, which is small compared to $\tau$. A small positive value $\delta$ results from a finite internal response time of the system. For each solution, an unique value $\eta$ can be chosen for which the oscillatory dynamics  becomes periodic with the period $\eta$.  Space-time plot corresponding to one impulse consists of a single localized perturbation surrounded by a plateaus of the almost constant amplitude (given by the quiescent regime) [Fig.~\ref{fig1}~(c)~,~(d)].  This structure is is topologically equivalent to dissipative solitons observed in spatially extended systems. The soliton shown in Fig. \ref{fig1} (c) corresponds to sharp decreasing of the signal $y(t)$ while the soliton depicted in Fig. \ref{fig1} (d) represents the increasing of the instantaneous value $y(t)$. Hence, one can distinguish dark [Fig \ref{fig1} (c)] and bright [Fig. \ref{fig1} (d)] solitons respectively \cite{kivshar2003}. Due to complete symmetry, each of them is obtained in Eq. (\ref{system}) with equal probability in a case of random initial conditions. Two quiescent states corresponding to the equilibrium points 1 and 2 are also stable and thus coexist with both the dark and bright solitons. 

Dynamics of the model (\ref{system}) is highly multistable. Beyond the two single solitons one can obtain (by applying different initial  conditions) a larger number of multi-soliton solutions, see examples in Fig. \ref{fig1} (e),(f), or ''soliton clusters'' as in Fig. \ref{fig1} (g)-(h). It should be noticed that soliton structures shown in the figures are obtained from random initial conditions. They hold the shape and do not decay with time preserving also the distance between them. There is, however, extremely small random drift of distances between solitons, but it does not play a principal role and the mean values remain approximately constant for long times of the simulations, up to $10^{6}$ delay time units  $n$ and more. In order to verify the soliton robustness we constructed a physical experiment, where the impact of fluctuations and the imperfectness of the setup is taken into account. 

%
\begin{figure}[t!]
\centering
\includegraphics[width=0.45\textwidth]{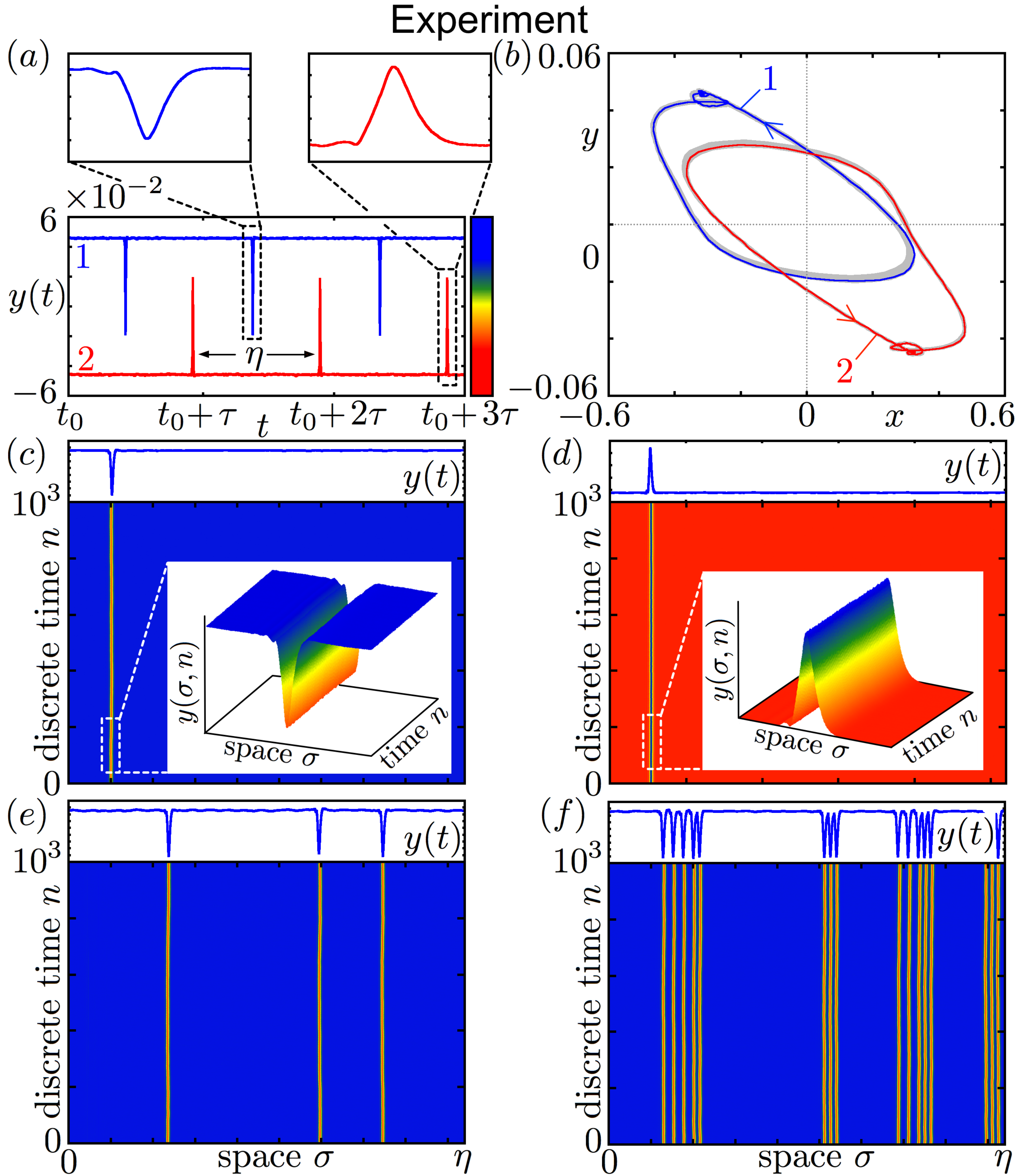} 
\caption{Experimental evidence of dissipative solitons in electronic setup. (a) Two realizations of $y(t)$ in time range $t\in[t_{0}:t_{0}+3\tau]$, shown in blue and red. (b) Corresponding projection of trajectories on phase plane ($x,y$). Grey background shows realizations at $t \in [0:60]$ sec. (c)-(f) Space-time plots of $y(t)$ corresponding to:  (c) one dark soliton (blue trajectory in panels (a),(b)); (d) one bright soliton (red trajectory in panels (a),(b)); (e) three dark solitons; (g) four bursts of totally sixteen dark solitons. Parameters of the setup (\ref{exp_setup}): $\varepsilon=0.1R_{x}/R_{y}=0.005$, $a_{e}=10.92$, $b_{e}=0.38$, $g=0.1$, $m=7$, $\tau=0.047$ sec. Quasi-space parameter is set to: $\eta=0.047194$ (panel (c)), $\eta=0.047217$ (panel (d)), $\eta=0.0482032$ (panel (e)), $\eta=0.0521999$ (panel~(f)).}
\label{fig2}
\end{figure}  

\subsection{Experimental realization}
Soliton structures which were observed in numerical simulations of the model (\ref{system}) have been also obtained in electronic experiment,  as illustrated in Fig. \ref{fig2}.  Space-time representation of the experimentally registered sequence of alternating impulses [Fig. \ref{fig2} (a)] allows to reveal the soliton patterns shown in Fig. \ref{fig2} (c)-(f). In the phase plane ($x$, $y$) the resulting dynamics of an experimental setup includes two quiescent states and two oscillatory loops corresponding to the impulses [Fig. \ref{fig2}~(b)]. Using temporal external excitation on the setup, one can obtain a single dark [Fig. \ref{fig2} (c)] or bright [Fig. \ref{fig2}~(d)] soliton, a number of the solitons [Fig. \ref{fig2} (e)],  as well as the soliton "bursts" [Fig. \ref{fig2} (f)]. In contrast to numerical simulations performed for the completely  symmetric case, where dark or bright soliton structures have equal probability, the experimental implementation of bright solitons appears to be much more difficult. These structures are very rare in our experiment. Such dissimilarity occurs because of some intrinsic features, imperfections and asymmetry of the setup, which will be discussed below in Ch. IV. 

%
\begin{figure}[t]
\centering
\includegraphics[width=0.45\textwidth]{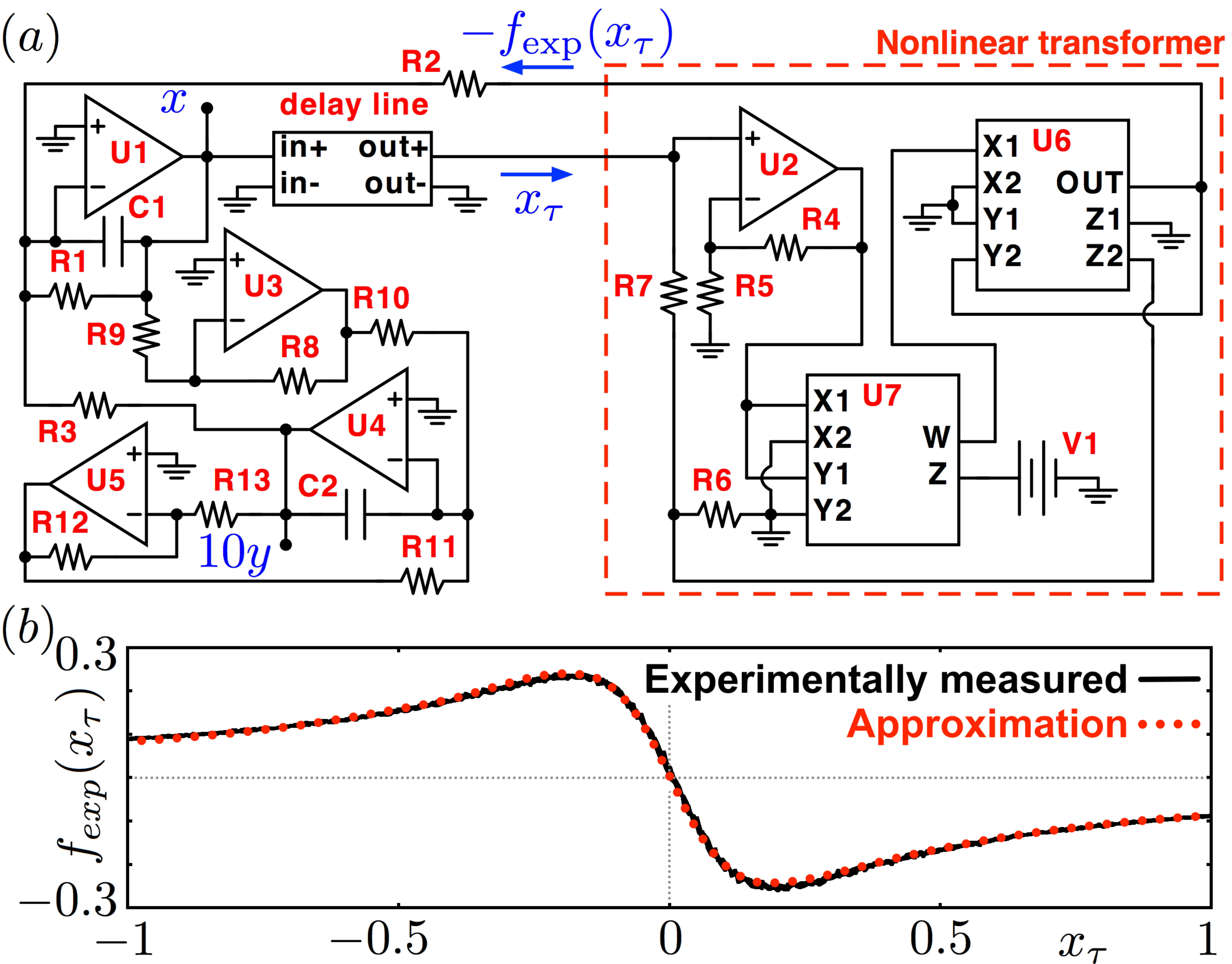} 
\caption{(a) Circuit diagram of experimental setup. Operational amplifiers U1-U5 are TL072CP, U6 is the analog multiplier AD534LD, U7 is the analog multiplier AD633JN, V1 is a source the DC voltage $V_{1}=10$ mV. Other elements are capacitors C1=C2=$50$ nF and resistors R1=R3=R5=R6=R9=R13=1 k$\Omega$, R2=2 k$\Omega$, R4=R7=9 k$\Omega$, R8=10 k$\Omega$, R10=R11=20 k$\Omega$, $6$ k$\Omega$ $\le$ R12 $\le8.5$ k$\Omega$. (b) Experimentally measured characteristics being responsible for nonlinear delayed feedback $f[x_{\tau}]$ (black solid line) and its approximation by function $f[x_{\tau}]=-x_{\tau}/(a_{e}x_{\tau}^2+b_{e})$ with $a_{e}=10.92$ and $b_{e}=0.38$ (red dotted line).}
\label{fig3}
\end{figure}  

Physical experiment was carried out with using an experimental prototype being an electronic model of the system (\ref{system}) implemented by principles of analog modelling \cite{luchinsky1998}. Detailed circuit diagram of the setup is shown in Fig. \ref{fig3} (a). It contains two integrators, U1 and U4, whose output voltages are taken as the dynamical variables, $x$ and $10y$, respectively. The experimental facility includes a time-delay line, which was realized using personal computer complemented by an acquisition board (National Instruments NI-PCI 6251). The time delay value is constant, $\tau=47$ ms. Equations describing operation of the experimental setup are the  following:
\begin{equation}
\label{exp_setup}
\left\lbrace
\begin{array}{l}
0.1R_{x}C \dfrac{dx}{dt} = -y-gx+\dfrac{1}{20}f[x_{\tau}],\\
R_{y}C\dfrac{dy}{dt}=x+my, \\
\end{array}
\right.
\end{equation}
where $x_{\tau}=x(t-\tau)$, $g=0.1$, $m=R12/R13$, $C=C_{1}=C_{2}=50$~nF,  $R_{x}=$1~K$\Omega$ is the resistance at the integrator U1 ($R_{1}=0.5 R_{2}=R_{3}=R_{x}=1~$K$\Omega$), $R_{y}=20$~K$\Omega$ is the resistance at the integrator U4 ($R_{10}=R_{11}=R_{y}=20$~K$\Omega$). Equations (\ref{exp_setup}) include the function $f[x_{\tau}]$ depending on delayed signal $x_{\tau}$ and being responsible for delayed feedback. The function $f[x_{\tau}]$ was realized by using a block of nonlinear transformation (the right part of the circuit depicted in Fig. \ref{fig3} (a)). The experimentally measured dependence $f[x_{\tau}]$ is presented in Fig. \ref{fig3} (b) (the black curve). The dependence $f[x_{\tau}]$ can be approximated by the formula $f[x_{\tau}]=-x_{\tau}/(a_{e}x_{\tau}^2+b_{e})$ with $a_{e}=10.92$ and $b_{e}=0.38$ (see the red dotted line in Fig. \ref{fig3} (b)). Then the quations of the experimental setup can be transformed into dimensionless system (\ref{system}) with $\varepsilon=\dfrac{0.1R_{x}}{R_{y}}$ by using substitution $t=t/\tau_0$  ($\tau_0=R_{y}C=10$~ms is the circuit's time constant) and new dynamical variables $x/V_{0}$ and $y/V_{0}$, where $V_{0}$ is the unity voltage, $V_{0}=1$~V.

Results of electronic experiment correspond to numerical simulation. Some quantitative difference between  phase trajectories registered numerically and experimentally (compare Fig.\ref{fig1} (b) and Fig.\ref{fig2} (b)) is due to the fact that equations (\ref{exp_setup}) were derived using standard approximation on operation amplifiers, which is common in electronics. This approach does not take into account features of  functioning of real operational amplifiers and analog multipliers and their distinctions from ideal ones. Despite of slight imperfection of the setup,  observed soliton structures in Fig. \ref{fig2} are robust and qualitatively correspond to the simulation. Individual solitons or soliton "bursts" obtained after external perturbation and short transient time always persist for all time of experiment running, up to several hours and more.

\section{Route from dissipative solitons to spatio-temporal chaos}

\begin{figure}[t!]
\centering
\includegraphics[width=0.44\textwidth]{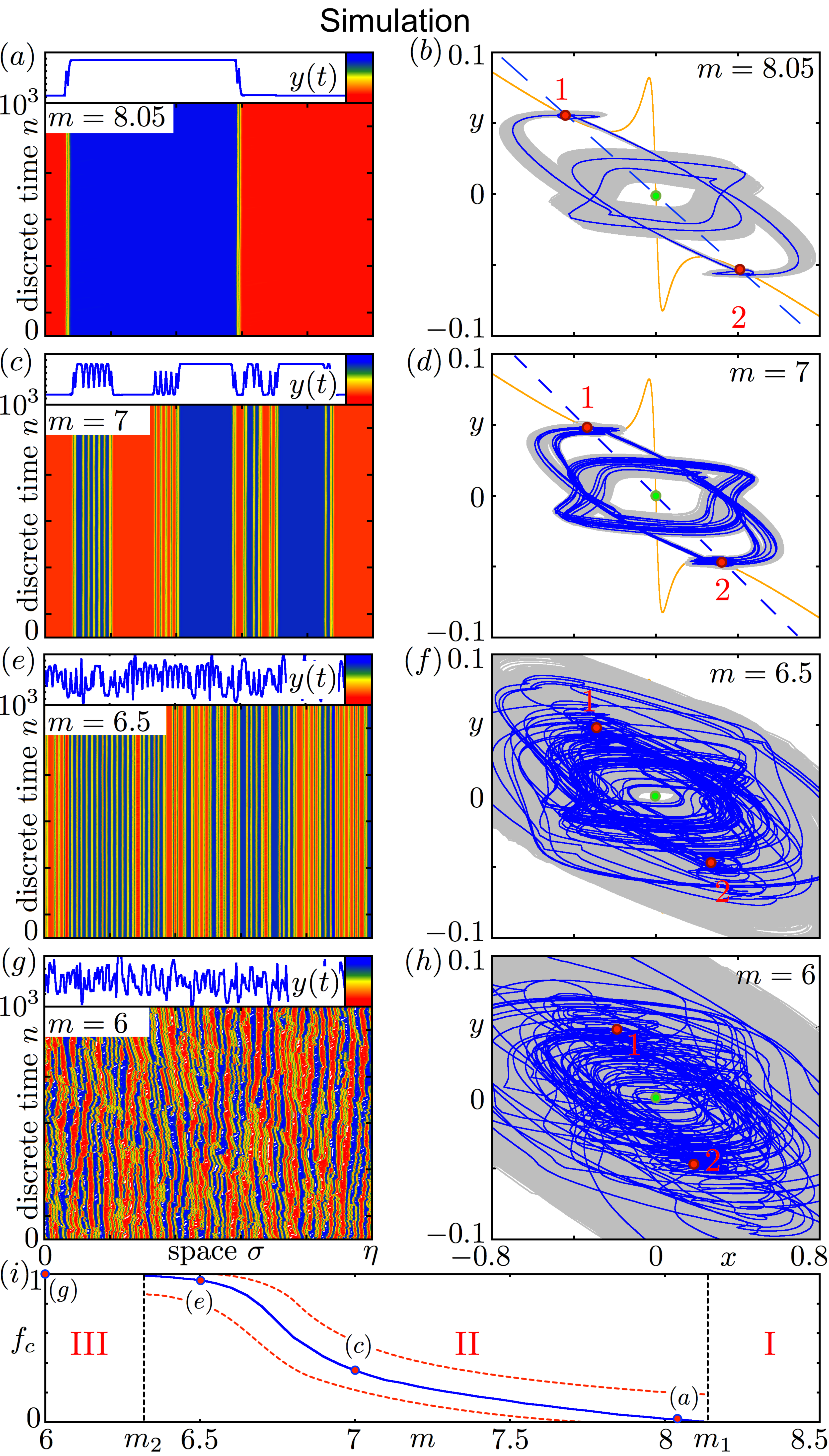} 
\caption{(a)-(h) Compound soliton-based structures  in symmetric model (\ref{system}) for different $m$. Space-time plots (left panels) and corresponding phase portraits (right panels): two-quiescent regime (panels (a),(b)), multiple soliton structures (panels (c)-(f)), spatio-temporal chaos (panels (g),(h)).  (i) Turbulent fraction $f_{c}$  (blue solid line) of the solution oscillatory activity versus $m$ built on 20 trails with random initial conditions;  maximal and minimal $f_{c}$ values obtained for each $m$ are shown by red dashed lines. Parameters are $\varepsilon=0.005$, $a=200$, $b=0.2$, $g=0.1$, $\tau=50$. Quasi-space parameters: $\eta=50.2359$ (a), $\eta=50.2298$ (c), $\eta=50.2283$ (e), $\eta=50.24$~(g).}
\label{fig4}
\end{figure}

Besides single and multiple soliton structures, symmetric model (\ref{system}) demonstrates other, compound states. An interesting regime (obtained as all others from random initial conditions) is shown in Fig. \ref{fig4} (a),(b). Here, the trajectory alternates between two quiescent steady states given by the fixed points 1 and 2.  It spends approximately equal time (around a half of the delay) close to each of the equilibria states, and there are no solitons.  Such kind structures were referred recently, as well as conditions of their stability, for a bistable oscillator with linear positive delayed feedback \cite{marino2014,marino2017}. Another example, of more involved behaviour is shown in Fig.\ref{fig4} (c),(d). There are both dark and bright solitons as well as their bursts intermingling with upper (shown red) and down (blue) quiescent intervals.

Choosing $m$ as a control parameter, we observe that if $8.5 \ge m \ge 8.1$ the system exhibits only two coexisting quiescent steady state regimes given by fixed points 1 and 2 (see Fig. \ref{fig1}~(b)).  When $m$ is less than $m_{1}=8.1$, soliton solutions arise together with a diversity of compound regimes like those two mentioned above [Fig. \ref{fig4}~(a)-(d)].  With further decrease of the parameter $m$, soliton structures become more profound, nevertheless preserving the periodicity in time with some period $\eta$ as illustrated in Fig. \ref{fig4}~(c),(d).  
Eventually, approaching the critical parameter value $m_2=6.3$ we observe complete soliton-filling of the virtual quasi-space [Fig. \ref{fig4}~(e),(f)].  Below $m_2$, the soliton dynamics ceases to exist giving rise to developed space-temporal chaos  [Fig. \ref{fig4}~(g)-(h)]. 

The described evolution of the dynamics can be characterized  in terms of {\it turbulent fraction} $f_c$ defined as a ratio of the intervals corresponding to oscillatory activity to the total interval length $\eta$.  A graph of the turbulent fraction $f_{c}$ for the system (\ref{system}) is shown in Fig. \ref{fig4} (i) as a blue curve. It was obtained by averaging of 20 respective values obtained from random initial conditions, calculated at each fixed value $m$ ($\Delta m=0.05$). As it can be seen,  $f_{c}$ equals to zero in the area of quiescent regimes (area I in Fig. \ref{fig4} (i)), and it gradually increases with decrease of $m$ reaching eventually the full value 1 at $m=m_2$  (along area II in Fig. \ref{fig4} (i) where the soliton-type structures are observed).  Below $m_2$, the quantity $f_{c}$ equals to 1 which corresponds to developed spatio-temporal chaos (area III).  An essential characteristic of the area II is multistability. Fixing parameter $m$ and starting from random initial conditions  one can obtain a variety of regimes with different number of the solitons (dark and bright) and alternations between quiescent steady states, charactered therefore by different values of the turbulent fraction $f_c$.  Upper and lower bounds of $f_{c}$ which were obtained in the 20 trial simulations are marked by red dashed curves in Fig. \ref{fig4} (i),  showing the maximal and minimal  possible values of $f_c$ depending on the parameter $m$.

%

Boundaries for the soliton area II are estimated in physical experiment as $m\in [5.77:8.3]$.  If $m>8.3$ the setup exhibits only two coexisting quiescent steady state regimes, similar to those in the model (\ref{system}) for $m>8.1$.  At the other side of the interval II, at $m=5.77$, the soliton dynamics collapses.  However, this transition is different from the model and is caused apparently by the changing the experimental circuit characteristics.  Indeed, for $m<5.77$ the experimental facility demonstrates an irregular regime which is not  associated with the behaviour of the model~(\ref{system}). 

\section{Role of asymmetry. Coarsening}

\begin{figure}[t!]
\centering
\includegraphics[width=0.45\textwidth]{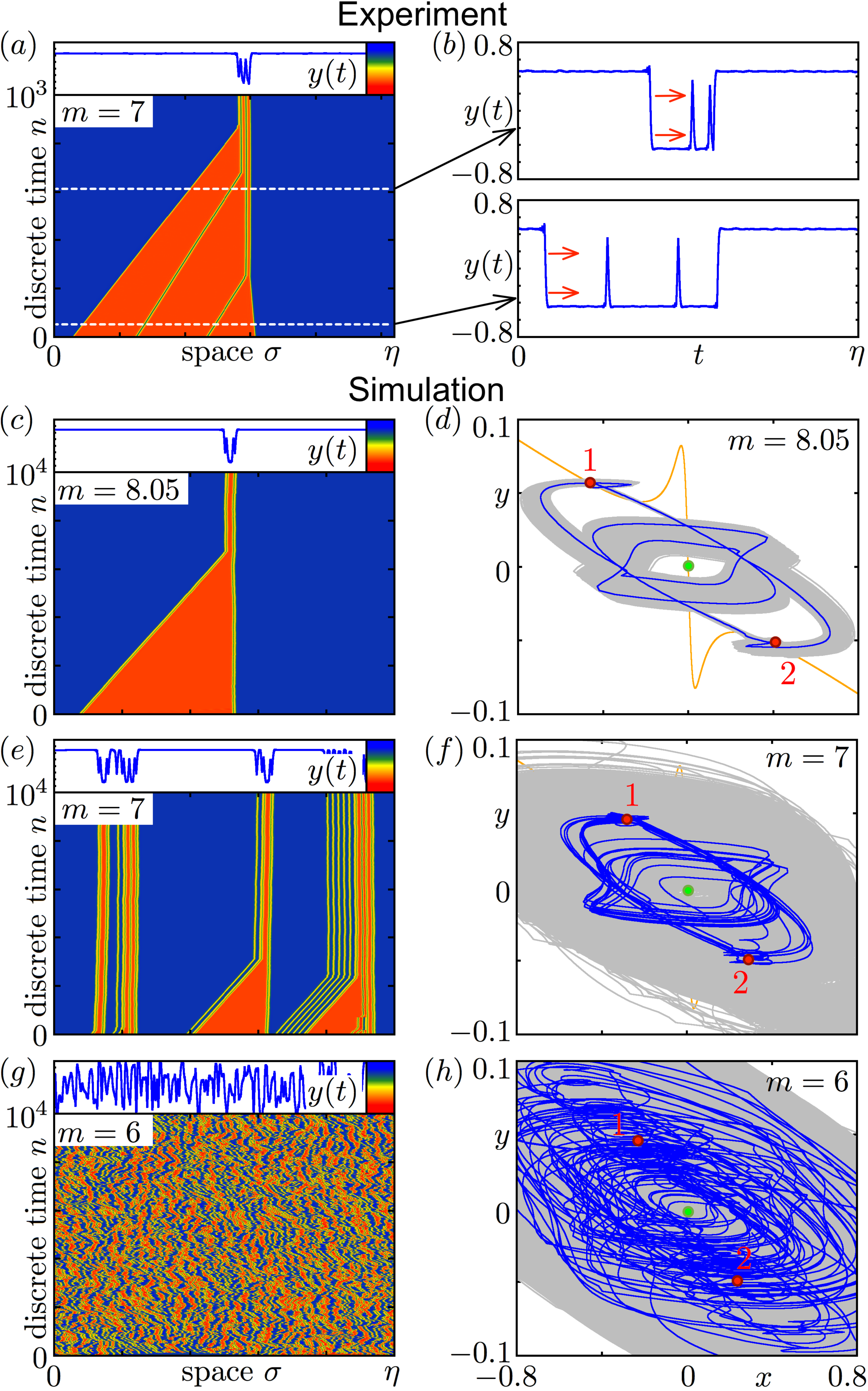} 
\caption{Effect of coarsening in system with asymmetry.  (a) Experimentally obtained space-time plot of $y(t)$  illustrating transitions from two (upper and down)  to one (upper) quiescent states; (b) Snapshots of $y(t)$ corresponding to different values of discrete time $n$ from panel (a). Parameters of the setup are the same as in Fig. \ref{fig2} (c)-(f) Examples of coarsening behaviour in model system (\ref{system}) with asymmetric  nonlinear function $f(x)=-\dfrac{x}{ax^2+b}-0.0002$.  (g)-(h) Chaotic behaviour beyond the soliton region ($m=6$).  System parameters are the same as in Fig. \ref{fig4} and Fig. \ref{fig2}. Quasi-space parameter $\eta$ equals to  $0.052243$ (panel (a)), $50.2282$ (panel (c)), $50.2283$ (panel (e)), $50.24$ (panel (g)).}
\label{fig5}
\end{figure}

Fig. \ref{fig2} (c),(d) illustrates two coexisting solitons, dark and bright,  obtained experimentally. They are slightly different, which is due to inevitable setup asymmetry (compare corresponding orbit loops 1 and 2 in Fig. \ref{fig1}~(b) for the symmetric model (1)).  As a consequence, one of the solitons (dark) is much more probable in the experimental trials as compared to the other (bright).  Moreover, we were not able to obtain also experimentally the regimes including alternations between the two quiescent states  (which are rather typical for completely symmetric model as illustrated in Fig. 4(a)-(d)).  These states, been generated at the beginning stage of the experiment, gradually disappear with time of the observation. This is the effect of coarsening \cite{giacomelli2012,yanchuk2017}: boundaries corresponding to transitions between the two quiescent states monotonically move in space and eventually, only dark soliton impulses survive [Fig. \ref{fig5} (a),(b)]. 

In order to derive the influence of asymmetry on the model dynamics, let us consider  Eqs. (\ref{system}) with a modified nonlinear function $f(x)=-\dfrac{x}{ax^2+b}-\xi$, where $\xi$ is a small parameter. Results of direct numerical simulation of the model (\ref{system}) with the asymmetric function $f$ are presented in Fig. \ref{fig5} (c)-(f), where the initial conditions are chosen as depicted in Fig. \ref{fig4} (a) and (c).  The effect of coarsening is clearly observed similar to the experimental setup [Fig.~\ref{fig5}~(a),(b)]. On the other hand,  the process of coarsening does not impact on the chaotic spatio-temporal dynamics observed in the area III [Fig.~\ref{fig5}~(g),(h)]. 

%


\section{Conclusions}

We have identified a novel mechanism for generation of single and multiple dissipative solitons in a second order modification of the Ikeda time-delayed equation including bistable nonlinear function with negative feedback. The study has been carried out by exploiting the correspondence of time-delayed equations with the spatially extended systems, where the solitons as well as other compound soliton-based structures become clearly visible. The reported results are supported by a physical electronic experiment which demonstrates an excellent qualitative agreement with the numerical simulation, proving in such a way the robustness and intrinsic system character of the phenomenon. This correspondence indicates a common, probably universal phenomenon in nonlinear delayed-feedback systems of very different nature.

The multiplicity of the dissipative solitons in the considered model can give a way to using them in the context of information storage. Indeed, by adjusting the initial conditions any number of solitons can be disposed in the virtual space and vice versa, each assigned soliton can be canceled by a slight external impact. There is, however, a restriction on the minimal distance between the units, which bounds their maximal possible number in the space.  The imposed restriction can be overcame by increasing of the time of delay. Then the information capacity of the system can fast grow. This intriguing issue could be a subject of future study.

\section{Acknowledgments}
This work was supported by Russian Ministry of Education and Science (project code 3.8616.2017/8.9).  We are grateful to Laurent Larger, Arkady Pikovsky, and  Serhiy Yanchuk for illuminating discussions. We also acknowledge support and hospitality of Institute FEMTO-ST of Besancon, Technical University of Berlin, and University of Potsdam.

\end{document}